# COUNTS-IN-CELLS ANALYSIS
# OF THE STATISTICAL DISTRIBUTION
# IN AN N-BODY SIMULATED UNIVERSE

Haruhiko Ueda

*Department of Astronomy, Kyoto University,
Kyoto 606-01, Japan, and*

*Uji Research Center
Yukawa Institute for Theoretical Physics
Kyoto University, Uji 611, Japan*

*E-mail ueda@yisun1.yukawa.kyoto-u.ac.jp*

and

Jun'ichi Yokoyama

*Uji Research Center
Yukawa Institute for Theoretical Physics
Kyoto University, Uji 611, Japan*

*E-mail yokoyama@yisun1.yukawa.kyoto-u.ac.jp*

## Abstract

Evolution of the statistical distribution of density field is investigated by means of a counts-in-cells method in a low-density cold-dark-matter simulated universe. Four theoretical distributions, i.e. the negative binomial distribution, the lognormal distribution, the Edgeworth series and the skewed lognormal distribution, are tested to fit the calculated distribution function, and it is shown that only the skewed lognormal distribution of second and third order can describe the evolution of the statistical distribution perfectly well from the initially Gaussian regime to the present stage. The effects of sparse sampling is also investigated and it is discussed that one should use a sample with number density of galaxies larger than $\sim 0.01 h^3 \mathrm{Mpc}^{-3}$ in order to recover underlying density distribution.

Key words: cosmology – galaxies: clustering

# 1 Introduction

It is widely believed today that the observed rich hierarchy of the large scale structure in the universe has emerged through gravitational instability of small initial inhomogeneity which is usually assumed to obey Gaussian statistics. While the statistical properties of Gaussian distribution are fully specified by the two-point correlation function, $\xi_2(r, t)$, or the power spectrum, even the perturbative evolution of fluctuations in the presence of gravity deforms the statistics as soon as second- and higher-order effects are taken into account (see e.g. Peebles 1980), not to mention the highly non-Gaussian nature of galaxy distribution today.

One way to quantify such deviations is to estimate higher-order reduced correlation functions, which are difficult to measure because they emerge only after carefully subtracting contributions of lower-order counterparts. Another quantity often used is the void probability, $P_0(V)$, namely the probability to find no galaxies in a volume $V$. This measure, however, only reflects the Poissonian nature of discrete distribution for small $V$ and it suffers from a large error for larger $V$ because the void probability becomes smaller and smaller as we increase $V$ (Gaztañaga and Yokoyama 1993). Thus we should not stick to the void probability but consider the probability to find $N$ galaxies, $P(N, V)$, as well. This quantity can easily be measured by the counts-in-cells method and it contains complete statistical information in the sense that it depends on arbitrary higher-order correlation functions averaged over the volume $V$ (White 1979).

Therefore it is an important issue of theoretical cosmology to clarify the nature of the statistical distributions of galaxies which sensible models of structure formation predict, so that one can compare various scenarios with observations and single out the correct model. As well important in understanding evolution of the universe is to trace time evolution of the statistical distribution, from initially Gaussian state to highly non-Gaussian distribution today.

In the present paper, using the results of $N$-body simulations, we analyze various proposals of theoretical modeling of statistical distribution and phenomenologically investigate which model fits the simulation best. There have been proposed a number of models of probability distribution function (PDF). The oldest one is probably the lognormal distribution which was first applied to galaxy distribution by Hubble (1934). The negative binomial distribution, which has the hierarchical property in higher-order reduced moments, has also been adopted by a number of authors (Fry 1986, Carruthers 1991, Gaztañaga & Yokoyama 1993, Bouchet et al. 1993). More recently the Edgeworth expansion around the Gaussian distribution has been proposed to modify it to incorporate higher-order correlations (Juszkiewicz et al. 1993). Similar expansion has also been applied to the lognormal distribution and called the skewed lognormal approximation (Colombi 1994). We compare these models with the results of counts analysis of simulated data at various epochs. There are, of course, other models of PDF, such as Saslaw's thermodynamic model (Saslaw and Hamilton 1984) or Balian-Schafer's model (1989).

We do not consider them here, which have been analyzed rather extensively already (Itoh, Inagaki, & Saslaw 1988, Suto, Itoh, & Inagaki 1990, Colombi et al. 1994).

The rest of the paper is organized as follows. In §2 we summarize properties of theoretical distribution we use. Those of simulated data and procedure of the analysis are described in §3. The results of counts analysis and comparison with theoretical models are given in §4. In §5 dependence of PDF on number density of the sample is reported. Finally §6 is devoted to discussion and conclusion.

# 2 Theoretical Models of statistical distribution

Here we summarize properties of four theoretical distributions we consider in turn.

## 2.1 *Negative Binomial Distribution*

Negative binomial (or modified Bose-Einstein) distribution has been used in a number of fields with different physical backgrounds such as quantum optics (Klauder and Sudarshan 1968), hadronic multiplicity (Carruthers and Shih 1983), galaxy counts in a Zwicky cluster (Carruthers and Minh 1983) in addition to the analysis of large scale galaxy counts (Fry 1986, Carruthers 1991, Gaztañaga & Yokoyama 1993, Bouchet et al. 1993). The distribution has been theoretically re-derived by Elizalde and Gaztañaga (1992) in an appropriate manner to our problem. That is, this distribution is obtained by incorporating two body correlation to the Poisson distribution in the simplest manner. The probability to find $N$ particles in a cell of volume $V$ is given by

$$P(N, V) = \frac{\overline{N}^N}{N!} (1 + \overline{\xi}_2 \overline{N})^{-N - \overline{\xi}_2^{-1}} \prod_{i=1}^{N-1} (1 + i \overline{\xi}_2), \qquad (1)$$

where $\overline{N} = \overline{N}(V)$ is the average number of particles contained in a cell and

$$\overline{\xi}_J(V) \equiv \frac{1}{V^N} \int_V \cdots \int_V \xi(\mathbf{x}_1, \mathbf{x}_2, \cdots \mathbf{x}_N) d\mathbf{x}_1 d\mathbf{x}_2 \cdots d\mathbf{x}_J, \qquad (2)$$

is the amplitude of the $J$-point correlation function averaged over the volume $V$.

One can soon notice that if we take $\overline{\xi}_2 \to 0$, negative binomial model reduces to the Poisson distribution,

$$P(N, V) = e^{-\overline{N}} \frac{\overline{N}^N}{N!}. \qquad (3)$$

In the continuum limit $\overline{N} \longrightarrow \infty$ with $x \equiv N/\overline{N}$ constant, the PDF yields

$$P(x) = \lim \overline{N} P(N, V) = \frac{1}{\Gamma(\overline{\xi}_2^{-1}) \overline{\xi}_2} (\overline{\xi}_2^{-1} x)^{\overline{\xi}_2^{-1} - 1} e^{-\overline{\xi}_2^{-1} x}. \qquad (4)$$

If we further take the limit $\overline{\xi}_2 \to 0$, it approaches Gaussian:

$$P(x) = \frac{1}{\sqrt{2\pi\overline{\xi}_2}} \exp\left[-\frac{(x-1)^2}{2\overline{\xi}_2}\right], \qquad \overline{\xi}_2 \ll 1. \tag{5}$$

One of the most important features of negative binomial model is that it obeys a hierarchical form of correlation functions. That is, the $J$-point reduced correlation function averaged over volume $V$ can be expressed as

$$\overline{\xi}_J(V) = S_J \left[\overline{\xi}_2(V)\right]^{J-1}. \tag{6}$$

This relation is observationally supported at least for $J = 3$ and 4 (Peebles 1980, Gaztañaga 1992), although such higher order correlation functions suffer from large error. For example, Gaztañaga (1992) finds

$$S_3 = 1.86 \pm 0.07 \quad (2.01 \pm 0.13),$$
$$S_4 = 4.15 \pm 0.60 \quad (4.96 \pm 0.88).$$

analyzing north CfAI (SSRS) data, respectively. In negative binomial distribution the coefficients are given by

$$S_J = (J-1)!, \tag{7}$$

for both discrete (1) and continuous (4) cases in close agreement with the observational analysis by Gaztañaga. Thus this model might describe the entire evolution of PDF in the universe, from initial Gaussian regime to present hierarchical regime.

## 2.2 *Lognormal Distribution*

Another model we consider is the lognormal distribution, which is related with a normal distribution

$$\phi(y) = \frac{1}{\sqrt{2\pi}\sigma_y} \exp\left[-\frac{(y-\overline{y})^2}{2\sigma_y^2}\right], \tag{8}$$

where $\overline{y}$, $\sigma_y$ are mean value and standard variation of $y$, respectively. Substituting $y$ by $\ln\rho$ with $\rho \equiv N/\overline{N}$ in the above distribution, we find a lognormal distribution function

$$P(N,V) = \frac{1}{N\sqrt{2\pi\ln(1+\overline{\xi}_2)}} \exp\left[-\frac{1}{2\ln(1+\overline{\xi}_2)}\left\{\ln\left(\frac{N\sqrt{1+\overline{\xi}_2}}{\overline{N}}\right)\right\}^2\right], \tag{9}$$

in the continuum limit. It also approaches Gaussian in the limit $\overline{\xi}_2 \longrightarrow 0$ where $\ln(1+\overline{\xi}_2) \longrightarrow \overline{\xi}_2$. In this distribution, the $n$-th order moment of counts is is expressed as

$$\langle N^n \rangle = \overline{N}^n (1+\overline{\xi})^{\frac{1}{2}n^2 - \frac{1}{2}n}. \tag{10}$$

Therefore if lognormal model represents matter distribution correctly, averaged two-point correlation function should satisfy the following relation,

$$\overline{\xi}_2 = \frac{\langle N^2 \rangle}{\overline{N}^2} - 1 = \left(\frac{\langle N^3 \rangle}{\overline{N}^3}\right)^{\frac{1}{3}} - 1 = \left(\frac{\langle N^4 \rangle}{\overline{N}^4}\right)^{\frac{1}{6}} - 1. \tag{11}$$

If an average cell does not contain a large enough number of galaxies, we must take discreteness effect into account. The PDF in such a case is obtained by the formula (White 1979; Fry 1985),

$$P(N, V) = \frac{1}{N!} \frac{d^N}{dt^N} M(V, t) \Big|_{t=-1}, \tag{12}$$

Here $M(V, t) \equiv \langle e^{Nt} \rangle$ is the moment generating function for a given volume $V$, which is given for the lognormal distribution by

$$M(V, t) = \int_0^\infty \frac{e^{xt}}{\sqrt{2\pi \ln(1 + \overline{\xi}_2)}} \exp\left\{ -\frac{[\ln x - \ln(\overline{N}/\sqrt{1 + \overline{\xi}_2})]^2}{2\ln(1 + \overline{\xi}_2)} \right\} \frac{dx}{x}. \tag{13}$$

We therefore obtain

$$P(N, V) = \frac{1}{N!} \int_{-\infty}^\infty \frac{dz}{\sqrt{2\pi \ln(1 + \overline{\xi}_2)}} \exp\left\{ Nz - e^z - \frac{[z - \ln(\overline{N}/\sqrt{1 + \overline{\xi}_2})]^2}{2\ln(1 + \overline{\xi}_2)} \right\}. \tag{14}$$

The lognormal matter distribution is obtained from the continuity equation in the nonlinear regime but with linear or Gaussian velocity fluctuations (Coles & Jones 1991). Thus it might be adopted as the preliminary model to describe general tendency of clustering in the weakly-nonlinear regime. Kofman et al. (1994) has shown that the lognormal distribution fits the PDF of $N-$body cold-dark-matter (CDM) simulation quite well but Bernardeau and Kofman (1994) discussed its successful fit just a coincidence due to the particular shape of the CDM power spectrum. See also Coles et al. (1993). On the other hand, Bouchet et al. (1993) applied it to IRAS galaxy redshift survey and concluded that both the lognormal and the negative binomial distributions fit the observed PDF well although the latter is somewhat better.

## 2.3 *Edgeworth series*

As mentioned in introduction, the primordial probability distribution function of density fluctuations is supposedly Gaussian, so the present PDF may be expressed by modifying Gaussian distribution to incorporate higher order correlations which are generated through gravitational clustering. In this way Juszkiewicz et al. (1993) proposed to apply the Edgeworth series to statistics of density fields. We consider PDF in terms of $\nu = \nu(N, V) \equiv \delta(N, V)/\sigma(V)$ where $\delta(N, V) \equiv$ $(N - \overline{N}(V))/\overline{N}(V)$ and $\sigma(V) \equiv \sqrt{\langle \delta^2(N, V) \rangle}$. Starting with the Gaussian distribution,

$$\phi(\nu) \equiv \frac{1}{\sqrt{2\pi}} \exp\left(-\frac{\nu^2}{2}\right), \tag{15}$$

we expand the PDF, $P(\nu)$, in terms of $\phi(\nu)$ and its derivatives as

$$P(\nu) = c_0 \, \phi(\nu) + \frac{c_1}{1!} \, \phi^{(1)}(\nu) + \frac{c_2}{2!} \, \phi^{(2)}(\nu) + \dots, \tag{16}$$

which is known as Gram-Charlier series (Cramér 1946). Here

$$\phi^{(\ell)}(\nu) \equiv \frac{d^\ell \phi}{d\nu^\ell} = (-1)^\ell \, H_\ell(\nu) \, \phi(\nu), \tag{17}$$

$$c_\ell = (-1)^\ell \int_{-\infty}^{\infty} H_\ell(\nu) \, P(\nu) \, d\nu, \tag{18}$$

and $H_\ell$ is the Hermite polynomial of degree $\ell$. For example, the lower-order coefficients are given by

$$c_0 = 1, \; c_1 = c_2 = 0, \; c_\ell = (-1)^\ell S_\ell \sigma^{\ell-2} \; (\text{for } 3 \leq \ell \leq 5), \; c_6 = S_6 \sigma^4 + S_3^2 \sigma^2, \tag{19}$$

with

$$S_3 \equiv \frac{\langle \nu^3 \rangle}{\sigma}, \qquad S_4 \equiv \frac{\langle \nu^4 \rangle - 3}{\sigma^2}. \tag{20}$$

The detailed expression of $S_5$ and $S_6$ are found in Juszkiewicz et al. (1993). Because $c_6$ in Gram-Charlier series have another $O(\sigma^2)$ contribution in addition to the $\phi^{(4)}$ term which is $O(\sigma^2)$, we have to rearrange the expansion by collecting all terms with the same powers of $\sigma$. Note that $S_\ell = O(\sigma^0)$ for all $\ell$ (Fry 1984; Bernardeau 1992). The result of such rearrangement is the so-called Edgeworth series in powers of $\sigma$. The zeroth- and the first-order Edgeworth expansion of PDF is simply $P(\nu) = \phi(\nu)$ because linear evolution does not alter the Gaussian shape. The second-order Edgeworth approximation reads

$$P(\nu) = \left[1 + \frac{1}{3!} S_3 \sigma H_3(\nu)\right] \phi(\nu), \tag{21}$$

and the third-order counterpart is given by

$$P(\nu) = \left[1 + \frac{1}{3!} S_3 \sigma H_3(\nu) + \frac{1}{4!} S_4 \sigma^2 H_4(\nu) + \frac{10}{6!} S_3^2 \sigma^2 H_6(\nu)\right] \phi(\nu). \tag{22}$$

Juszkiewicz et al. (1993) have shown that Edgeworth approximation fits the evolution of density fluctuations of $N$-body simulations with $n = -1$ power-law power spectrum evolved from Gaussian form for $\sigma \lesssim 1/4$.

## 2.4  *Skewed Lognormal Distribution*

Skewed lognormal approximation, which combines the lognormal distribution and the Edgeworth series, has been proposed by Colombi (1994). If we substitute $\nu$ by $v = v(N, V) \equiv \Phi(N, V)/\sigma_\Phi(V)$, with $\Phi(N, V) \equiv \ln \rho - \langle \ln \rho \rangle$, $\sigma_\Phi(V) \equiv \sqrt{\langle \Phi^2 \rangle}$ and $\rho \equiv N/\overline{N}$ in equation (15) and perform the same procedures, we obtain a skewed lognormal distribution. It is soon noticed that first-order skewed lognormal approximation

$$P(N, V) = \frac{\phi(v)}{N}, \qquad v = v(N, V) = \frac{\ln N - \langle \ln N \rangle}{\sqrt{\langle (\ln N - \langle \ln N \rangle)^2 \rangle}}, \qquad (23)$$

is nothing but the lognormal distribution. The second-order approximation reads

$$P(N, V) = \left[ 1 + \frac{1}{3!} T_3 \sigma_\Phi H_3(v) \right] \frac{\phi(v)}{N}, \qquad (24)$$

and the third-order one is given by

$$P(N, V) = \left[ 1 + \frac{1}{3!} T_3 \sigma_\Phi H_3(v) + \frac{1}{4!} T_4 \sigma_\Phi^2 H_4(v) + \frac{10}{6!} T_3^2 \sigma_\Phi^2 H_6(v) \right] \frac{\phi(v)}{N}. \qquad (25)$$

where

$$T_3(V) \equiv \frac{\langle (\ln \rho - \langle \ln \rho \rangle)^3 \rangle}{\sigma_\Phi^4}, \qquad (26)$$

$$T_4(V) \equiv \frac{\langle (\ln \rho - \langle \ln \rho \rangle)^4 \rangle - 3\sigma_\Phi^4}{\sigma_\Phi^6}. \qquad (27)$$

While this distribution also suffers from the same problem of positive non-definiteness as the Edgeworth series around the Gaussian distribution in principle, the former surpasses the latter in that the expansion parameter $\sigma_\Phi$ remains smaller than unity even when $\sigma$ is considerably larger than one, namely, in the nonlinear stage. In fact, using the $N$-body simulation of a CDM model, Colombi (1994) showed that this approximation also successfully describes the distribution function of density fluctuations in the evolved universe corresponding to the present.

# 3   Simulation Data and Counts Analysis

## 3.1   *N-body Simulation Data*

As a cosmological model we make use of the results of $N$-body simulation of a CDM model with low density parameter performed by Suginohara and Suto (1991) with a positive cosmological constant and primordially scale-invariant fluctuation spectrum. More specifically, its physical parameters are taken as follows: total particle number $N_{\text{tot}} = 64^3 = 262144$, the Hubble constant, $H_0 =$ $100h\,\mathrm{km\,s^{-1}\,Mpc^{-1}} = 100\,\mathrm{km\,s^{-1}\,Mpc^{-1}}$, present density parameter $\Omega_0 = 0.2$, and the cosmological constant, $\lambda \equiv \Lambda/(3H_0^2) = 0.8(= 1 - \Omega_0)$. The simulation was carried out with the hierarchical tree code in a cubic volume of $L_b^3$ with a periodic boundary condition. The comoving size of the simulation box corresponds to $L_b = 100\mathrm{Mpc}$ today, and the mass of each particle to $2 \times 10^{11} M_\odot$, roughly equal to a typical galactic mass. The amplitude of initial fluctuations is normalized so that the correlation function has the unit amplitude on scale $r \simeq 5h^{-1}\mathrm{Mpc}$ at the epoch corresponding to the present.

Detailed analysis of the simulation has shown that this model is among the most promising one to reproduce the observed large scale structure (Ueda et al. 1993) without nontrivial biasing as far as we normalize the amplitude of fluctuations using the correlation function. On the other hand, normalizing the amplitude in terms of COBE data, Efstathiou et al. (1992) concluded a substantial amount of anti-biasing is required for spatially-flat low-density CDM model with scale-invariant initial fluctuations. Since our simulation is sensitive to the comoving scale only up to $100h^{-1}\mathrm{Mpc}$ it could be reconciled with COBE normalization scheme without biasing if we would assume a different spectrum of initial fluctuations which has a bend on large scale.

In order to examine the evolution of the statistical distribution, we use position data of each particle at four different epoch at $a = 1.0, 3.0, 5.0,$ and $6.0$, where $a$ is the scale factor, which is normalized to unity at the initial epoch, when statistical distribution practically obeys Gaussian, and $a = 6.0$ corresponds to the present. Figure 1 depicts time evolution of $\bar{\xi}_2(V, t)$ as a function of the radius of a cell.

## 3.2 *Counts-in-Cells Analysis*

In performing counts-in-cells analysis, we adopt spherical cells. We randomly generate 100000 points in the simulation box, which serve as the center of each cell, and calculate distance between each point and each galaxy, paying attention to the periodic nature of the entire box. From these data we can easily reproduce the PDF, $P(N, V)$, for arbitrary volume with $L_b^3/64^3 = 3.8 \times 10^{-6} L_b^3 \ll V \ll L_b^3$. In practice we took $V$ in the range

$$2.7 \times 10^{-4} L_b^3 \lesssim V \lesssim 3.3 \times 10^{-2} L_b^3, \tag{28}$$

corresponding to the range of the radius of a cell,

$$0.04 L_b < r < 0.2 L_b. \tag{29}$$

As is seen above, we place a fixed number of cells independent of their volume. As long as this number is large enough, it does not affect the shape of the estimated PDF, although it does affect estimation of errors in the PDF reproduced, which we do not work out here.

# 4 Fitting evolution of the PDF

## 4.1 *Negative Binomial versus Lognormal distributions*

We shall now display the results of counts-in-cells analysis first in comparison with negative binomial and lognormal distributions. In figures 2 the histogram of calculated PDF is depicted for $r = 0.04L_b$, $0.08L_b$, and $0.2L_b$, together with negative binomial (solid line) and lognormal (dashed line) distributions with the same values of $\overline{N}$ and $\overline{\xi}_2$. As is seen there both models reproduce PDF reasonably well for $r = 0.2L_b$. For the cases $r = 0.04L_b$ and $0.08L_b$, however, although lognormal distribution is still applicable, significant deviation is observed between negative binomial model and the real PDF. Thus we conclude it is not an appropriate model for PDF in highly nonlinear regime.

As a different test of the lognormal distribution, let us examine lower-order moments of counts in terms of the predicted relation (11). Figures 3 depict $\overline{\xi}_2 = \langle N^2 \rangle / \overline{N}^2 - 1$ (solid line) as well as $(\langle N^3 \rangle / \overline{N}^3)^{1/3} - 1$ (dashed line) and $(\langle N^4 \rangle / \overline{N}^4)^{1/6} - 1$ (dot-dashed line) as a function of $r$ at various epochs. All of them should agree with each other if the underlying PDF obeys the lognormal distribution. As is seen there, these three lines practically coincide with each other at the initial epoch where particles are distributed using the Zel'dovich approximation. This result is consistent with the analysis of Kofman et al. (1994) where they found the PDF derived by the Zel'dovich approximation is very similar to the lognormal distribution in the quasi-linear regime. The agreement remains valid on large scales up to the epoch $a = 6.0$. On small scales with $\overline{\xi}_2 \gtrsim 1$, however, these lines starts to deviate from each other as the nonlinearity increases.

This suggests that the lognormal distribution, too, is not an adequate model to characterize the PDF in highly nonlinear regime and some correction should be taken into account. In the next subsection we study Edgeworth series around Gaussian distribution as a preparation of such improvement.

## 4.2 *Edgeworth expansion around Gaussian distribution*

One can consider the Edgeworth expansion introduced in §2.3 as a way to obtain a PDF which correctly reproduces arbitrary higher-order moments of the count. For example, if we adopt (22) it can reproduce both $\langle \nu^3 \rangle$ and $\langle \nu^4 \rangle$ as long as we insert correct observed values to $S_3$ and $S_4$. Figures 4 are the comparison between the calculated PDF and first- (solid), second- (dashed), third-order (dot-dashed line) Edgeworth series, respectively. The first-order one is nothing but the Gaussian distribution. Apparently, although the second- and third-order Edgeworth PDF fits the histogram quite well on large scale with $r = 0.2L_b$, significant deviation is observed in the later epochs on small scale $r = 0.04L_b$. As originally stressed by Juszkiewicz et al (1993), these PDFs are not appropriate to fit the observed PDF in highly nonlinear regime. In particular, the fact that they contain the Hermite polynomials implies that they become oscillatory and not positive definite with larger values of $\sigma$.

## 4.3 *Skewed Lognormal Distribution and Kolmogorov-Smirnov test*

Having confirmed the expansion around the Gaussian distribution does not provide a good approximation, we next consider that around the lognormal distribution following Colombi (1994). In figures 5, comparison between PDF derived from counts-in-cells analysis and skewed lognormal distribution with $r = 0.04L_b$, $0.08L_b$ and $0.2L_b$ is found. In each panel, solid, dashed, and dot-dashed lines represent first- (*i.e.* lognormal), second-, and third-order skewed lognormal distributions, respectively. In these figures, we find that the second- and the third-order distributions reproduce the PDF extremely well for all the cases, better than the lognormal model. In fact, the second- and the third-order distributions look degenerate in most cases.

In order to obtain more quantitative conclusion, let us adopt the Kolmogorov-Smirnov (KS) test using unbinned data. This test is applicable for cumulative distribution function (CDF) $C(N, V) \equiv \sum_{J=0}^{N} P(J, V)$. For comparing observed CDF, $C_{\rm obs}(N, V)$, with theoretical distributions, $C_{\rm th}(N, V)$, such as cumulative lognormal or cumulative skewed lognormal distributions, we introduce K-S static $D$ as

$$D \equiv \max_{N} |C_{\rm obs}(N, V) - C_{\rm th}(N, V)|, \tag{30}$$

which is the maximum value of the absolute difference between the observed CDF and the theoretical CDF. The significance level of an observed value of $D$ is given approximately by

$$P(D > \text{observed}) = Q_{KS}(\sqrt{T}D), \tag{31}$$

where

$$Q_{KS}(x) \equiv 2\sum_{j=1}^{\infty} (-)^{j-1} \exp(-2j^2 x^2). \tag{32}$$

Here $T$ is the maximum value of counts observed in a cell. Notice that $Q_{KS}(x)$ is a monotonic function with $Q_{KS}(0) = 1$ and $Q_{KS}(\infty) = 0$. Therefore if two CDF is identical to each other, $P(D > \text{observed})$ becomes unity, and its deviation from unity represents discrepancy between these CDFs. Table 1 shows the results of $P(D > \text{observed})$ between the observed PDF and $k$−th order skewed lognormal models. From this table, we can again see that the second- and third-order approximation are superior to the lognormal distribution function for which $P(D > \text{observed})$ is considerably smaller than unity for $r = 0.04L_b$ $a = 3.0, 5.0$, and $6.0$.

Thus the KS test shows the fact that the lognormal distribution function does not reproduce PDF correctly in highly nonlinear regime, although the deviation is not marked. On the other hand, on the scales of our interest the second- and the third-order approximation fit the data equally well. We therefore conclude that the second-order skewed lognormal distribution suffices to describe the evolution of the statistical distribution from its Gaussian initial condition to the present universe.

In addition to the KS test, we have estimated normalized skewness, $S$, and normalized kurtosis, $K$, of count in the skewed lognormal approximation, because, unlike in the Edgeworth series around the Gaussian distributions, these quantities are not guaranteed to match with the observed values automatically in this approximation. They are defined by

$$S \equiv \frac{\langle \delta^3(V) \rangle}{\langle \delta^2(V) \rangle}, \qquad K \equiv \frac{\langle \delta^4(V) \rangle - 3\langle \delta^2(V) \rangle^2}{\langle \delta^2(V) \rangle^3}, \qquad (33)$$

and the results are depicted in figures 6 where solid line represents the observed values and short dashed, dot-short dashed and dot-long dashed lines are the results calculated from the lognormal distribution, second- and third-order skewed lognormal models, respectively. These figures also show that skewed lognormal approximation is better than the lognormal distribution and that the second- and the third-order approximations are equally satisfactory.

# 5 Effects of sparse sampling

So far we have analyzed the statistical distribution using all of the 262144 particles in the simulation and we have fairly been free from the effects of discreteness. In the counts analysis of the actual universe, however, we must inevitably use volume-limited samples which contains only a limited and rather small number of galaxies. Here we analyze the dependence of PDF on the number, $N_{\rm obs}$, of total "galaxies" we "observe" in the simulation box at the present epoch $a = 6.0$. In order to simulate such a sparsely traced samples, we perform counts analysis using some limited portions of particles, which are selected randomly because no information is contained in the simulation on the luminosity of each particle.

Figures 7 depicts the result, where the solid line represents the negative binomial distribution, eq. (1), and dot-dashed line stands for the lognormal distribution with discreteness effects taken into account, eq. (14). For comparison, the lognormal model in the continuum limit, eq. (9), is also depicted in the figures with dashed line. As above, three different scales are probed corresponding to $r = 4, 8$ and $20h^{-1}$Mpc for which $\overline{\xi}_2$ are calculated to be $\overline{\xi}_2 = 2.9$, 0.90, and 0.18 respectively. As is seen there for $N_{\rm obs} \leq 3000$ both the negative binomial and the lognormal distributions fit the data equally well except for the low-count tail of the case with $(r, N_{\rm obs}) = (0.08L_b, 3000)$, but in some cases the negative binomial distribution seems to characterize the histogram somewhat better. On the other hand, as $N_{\rm obs}$ is increased significant deviation starts to appear between the negative binomial distribution and the calculated PDF especially in the low-count region on scales in the quasi-linear or semi-nonlinear regime.

Note again that presently available volume-limited samples of galaxies contain rather small numbers of galaxies. For example, $N_{\rm obs} = 700$ roughly corresponds to the number density of galaxies in the samples CfAN80 and SSRS80 used by Gaztañaga and Yokoyama (1993) and $N_{obs} = 3000$ to that in the IRAS sample of Bouchet et al. (1993) with radius $39h^{-1}\mathrm{Mpc}$. We can hardly distinguish the negative binomial and the lognormal distributions out of these samples. We need a catalog with larger number of galaxies.

# 6    Conclusion

In the present paper we have investigated evolution of statistical distribution in cosmological $N$-body self gravitating system. Using the results of a low-density CDM simulation, which reproduces various observational data well such as two-point correlation function of galaxies and that of clusters without nontrivial biasing, we have phenomenologically examined which theoretical model fits the simulated PDF best in terms of the counts-in-cells method. As for the theoretical distributions, we have considered four different approaches, namely, the negative binomial model, the lognormal distribution, Edgeworth series around Gaussian, and the skewed lognormal approximation, all of which are theoretically well motivated.

Our results show that the negative binomial model and Edgeworth series are not suitable to fully describe the statistical distribution of underlying density field. The failure of the latter in nonlinear regime is a good example of the fact that precise knowledge of lower-order moments is far from sufficient to specify the shape of the PDF.

On the other hand, we have found the lognormal distribution matches with the observed PDF reasonably well except in highly nonlinear regime. If we improve it to the skewed lognormal distribution, the agreement is extended to the nonlinear stage as well. The price we must pay for it is additional input parameters $T_3$, $T_4$, etc.. Our analysis, however, has shown that inclusion of only $T_3$ is sufficient to obtain a satisfactory approximation to the PDF at least up to the present stage of clustering. As emphasized by Colombi (1994), the skewed lognormal distribution is also doomed to failure in extremely nonlinear regime. Fortunately, however, as far as cosmological statistical distribution probed by galaxies are concerned, this would only take place in a distant future.

Armed with the fact that the underlying statistical distribution in the present model is satisfactorily characterized by the (skewed) lognormal distribution, we have also investigated the dependence of PDF on the number of galaxies we use to calculate it. We have shown that, for the range of $N_{obs}$ with comparable number density of galaxies to typical volume-limited samples available today, there exists at least one different distribution that reproduces the calculated PDF well but does not reflect the underlying matter distribution properly, namely, the negative binomial distribution. Thus the previous conclusions of the observational analyses (Gaztañaga & Yokoyama 1993; Bouchet et al. 1993) that this distribution fits the observed PDF might have nothing to do with the real statistical distribution of our universe. In fact, to single out the correct statistical distribution we should not only probe the PDF on various length scales with different values of $\overline{\xi}_2$ but also use volume-limited samples with large enough number density, $n_{\text{gal}}$, of galaxies, say, $n_{\text{gal}} \gtrsim 0.01 h^3 \text{Mpc}^{-1}$.

Acknowledgments

We thank Tatsushi Suginohara and Yasushi Suto for providing us their $N$-body simulation data and for valuable comments. The present analysis were carried out on SUN SPARC stations and Hitachi 3050RX at Educational Center for Information Processing, Kyoto University. This research was supported in part by Research Fellowship of the Japan Society for the Promotion of Science for Young Scientists No. 3004(HU) and by the Japanese Grant-in-Aid for Scientific Research Fund of Ministry of Education, Science, and Culture, No. 06740216(JY).

# REFERENCES

Balian, R., Schaeffer, R. 1989,  A&A,  220, 1

Bernardeau, F. 1992,  ApJ,  392, 1

Bernardeau, F., Kofman, L. 1994, submitted to ApJ

Bouchet, F.R., Strauss, M.A., Davis, M., Fisher, K.B., Yahil, A., Huchra, J. 1993,  ApJ, 417, 36

Carruthers, P., Minh, D.-V. 1983,  Phys. Lett. B,  131, 116

Carruthers, P., Shih, C.C. 1983,  Phys. Lett. B,  127, 242

Carruthers, P. 1991,   ApJ,   380, 24

Coles, P., Jones, B. 1991,  MNRAS,  248, 1

Coles, P., Melott, A., 1993,  MNRAS,  260, 765

Colombi, S. 1994,  ApJ,  435, 536

Colombi, S., Bouchet, F.R., Schaeffer, R. 1994, Fermilab preprint FERMILAB-Pub-94/229-A

Cramér 1946,  Mathematical Methods of Statistics (Princeton University Press: Princeton)

Elizalde, E., Gaztañaga, E. 1992,  MNRAS,  254, 247

Fry, J.N. 1984,  ApJ,  227, L5

Fry, J.N. 1985,  ApJ,  289, 10

Fry, J.N. 1986,  ApJ,  306, 358

Gaztañaga, E. 1992,  ApJ,  398, L17

Gaztañaga, E., Yokoyama, J. 1993,  ApJ,  403, 450

Hubble, E. 1934,  ApJ,  79, 8

Itoh, M., Inagaki, S., Saslaw, W.C. 1988,  ApJ,  331, 45

Juszkiewicz, R., Weinberg, D., Amsterdamski, P., Chodorowski, M., Bouchet, F.R. 1993, preprint

Klauder, J.R., Sudarshan, E.C.G. 1968,  Fundamentals of Quantum Optics ( Benjamin: New York)

Kofman, L., Bertschinger, E., Gelb, M.J., Nusser, A.,Dekel, A. 1994,  ApJ,  420, 44

Peebles P.J.E. 1980,   The Large Scale Structure of the Universe (Princeton University Press: Princeton)

Saslaw W.C., Hamilton A.J.S. 1984,  ApJ,  276, 13

Suginohara, T., Suto, Y. 1991,   PASJ,  43, L17

Suto, Y., Itoh, M., Inagaki, S. 1990,  ApJ,  350, 492

Ueda, H., Itoh, M., Suto, Y. 1993,  ApJ,  408, 3

White, S.D. 1979,  MNRAS,  186, 145

**Table 1. Result of Kolmogorov-Smirnov Test**

| $r$ | $0.04L_b$ | | | | $0.2L_b$ | | | |
|---|---|---|---|---|---|---|---|---|
| $a$ | 1.0 | 3.0 | 5.0 | 6.0 | 1.0 | 3.0 | 5.0 | 6.0 |
| k=1 | 1.00 | 0.97 | 0.56 | 0.29 | 1.00 | 1.00 | 1.00 | 1.00 |
| k=2 | 1.00 | 1.00 | 1.00 | 1.00 | 1.00 | 1.00 | 1.00 | 1.00 |
| k=3 | 1.00 | 1.00 | 1.00 | 1.00 | 1.00 | 1.00 | 1.00 | 1.00 |

KS statistic $P(D > \text{observed})$ for the $k-$th-order skewed lognormal distribution as a function of $r$ and $a$.

## Figure Captions

**Figure 1 :** Time evolution of $\overline{\xi}_2(V, t)$ as a function of the radius (in unit of $L_b$) of a cell.

**Figure 2 :** Time evolution of PDF in comparison with the negative binomial and the lognormal distributions. In each panel, histogram, solid and dashed lines represent PDF, negative binomial and lognormal distributions, respectively. Three different scales, $r = 0.04L_b, 0.08L_b$, and $0.2L_b$ are represented.

**Figure 3 :** Averaged two-point correlations function which are estimated from $\langle N^2 \rangle / \overline{N}^2 - 1$ (solid line), $(\langle N^3 \rangle / \overline{N}^3)^{1/3} - 1$ (doted line) and $(\langle N^4 \rangle / \overline{N}^4)^{1/6} - 1$ (dashed line).

**Figure 4 :** Comparison between PDF $P(\nu)$ and the Edgeworth series. Histogram represents PDF estimated from counts-in-cells analysis. Solid, dashed and dot-dashed lines represent first-, second- and third- order Edgeworth series.

**Figure 5 :** Comparison between PDF and the skewed lognormal distribution. In each panel, histogram represents calculated PDF and solid, dashed, and dot-dashed lines represent first-, second- and third-order skewed lognormal distributions. The second- and the third-order distributions are hardly distinguishable from each other.

**Figure 6 :** Skewness and kurtosis of density fluctuations as a function of $V$ in unit of $L_b^3$. Solid, dashed, dot-dashed and dot-long dashed lines represent skewness or kurtosis which are estimated from the actual PDF, first-, second- and third-order skewed lognormal distributions, respectively.

**Figure 7 :** Dependence of PDF on the numbers of galaxies, $N_{\rm obs}$, which are selected randomly from the simulation. In each panel, histogram, solid, dashed, and dot-dashed lines represent PDF, negative binomial, continuous lognormal, discrete lognormal models respectively. For the case $(r, N_{\rm obs}) = (0.04L_b, 700)$ solid and dot-dashed lines are degenerate, while for $(r, N_{\rm obs}) = (0.2L_b, 10000)$, $(0.2L_b, 30000)$ discrete and continuous lognormal distributions coincide with each other.

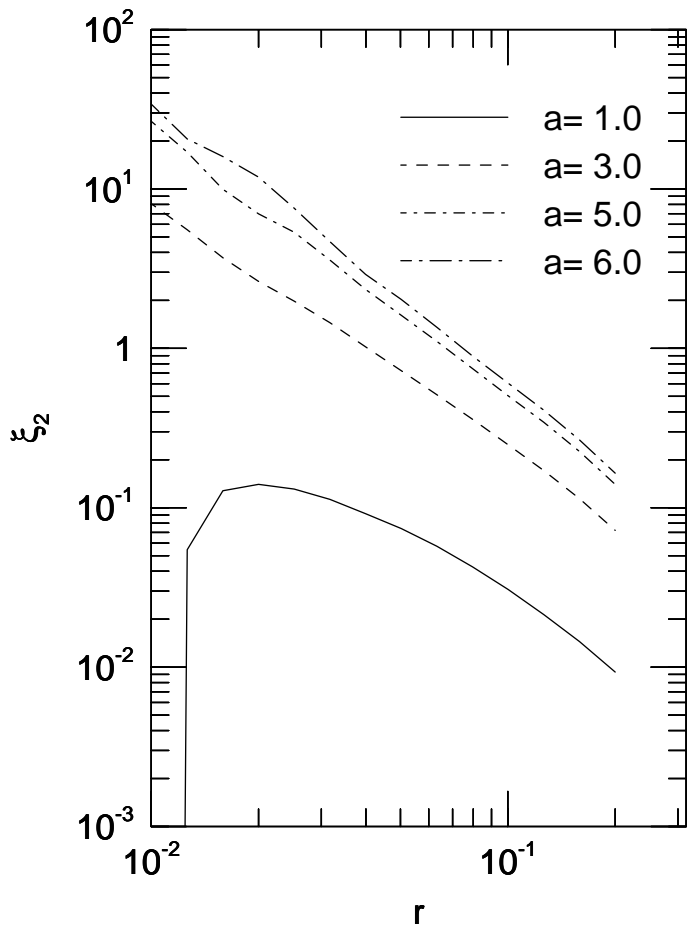

Figure 1

Figure 2

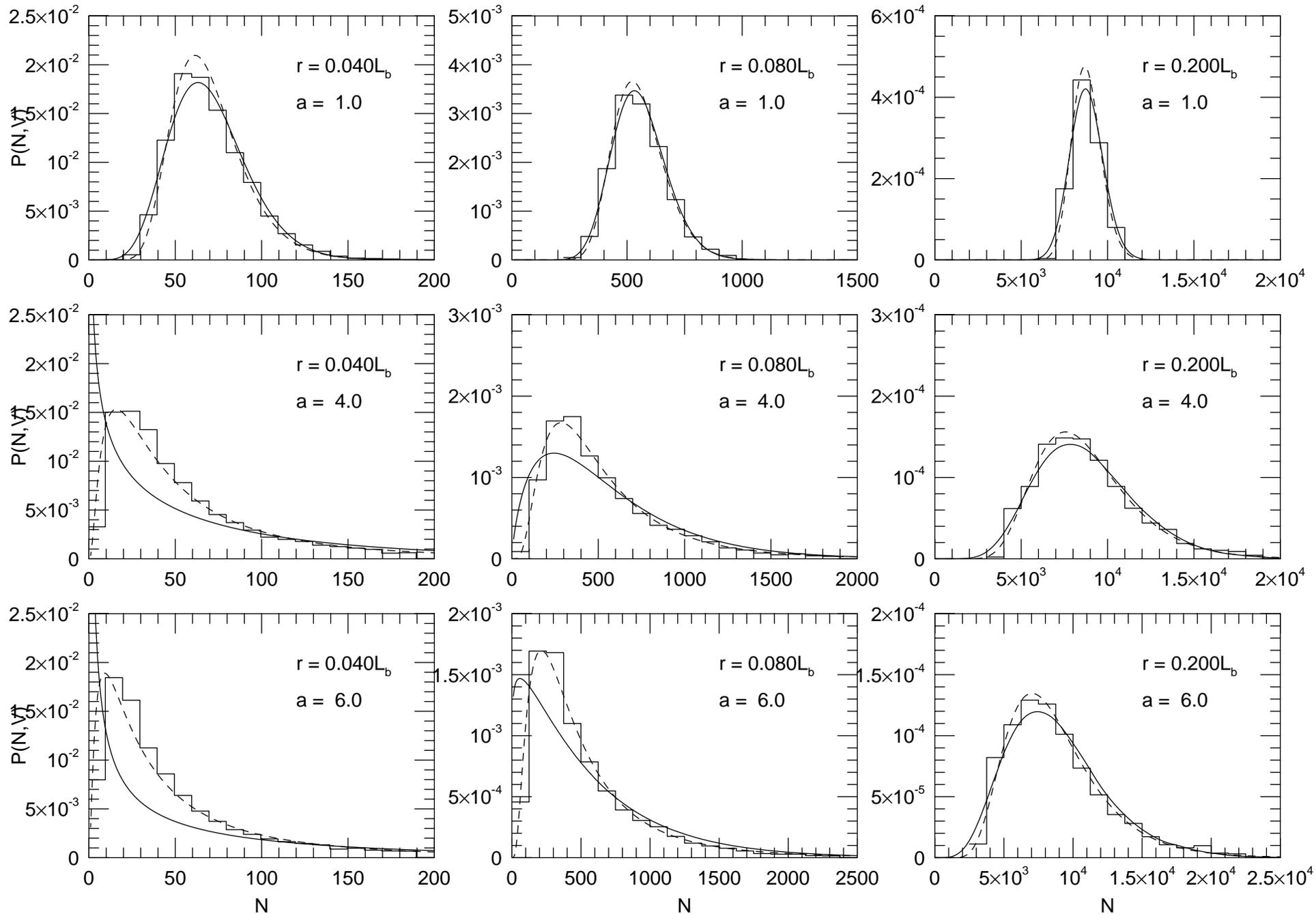

Figure 3

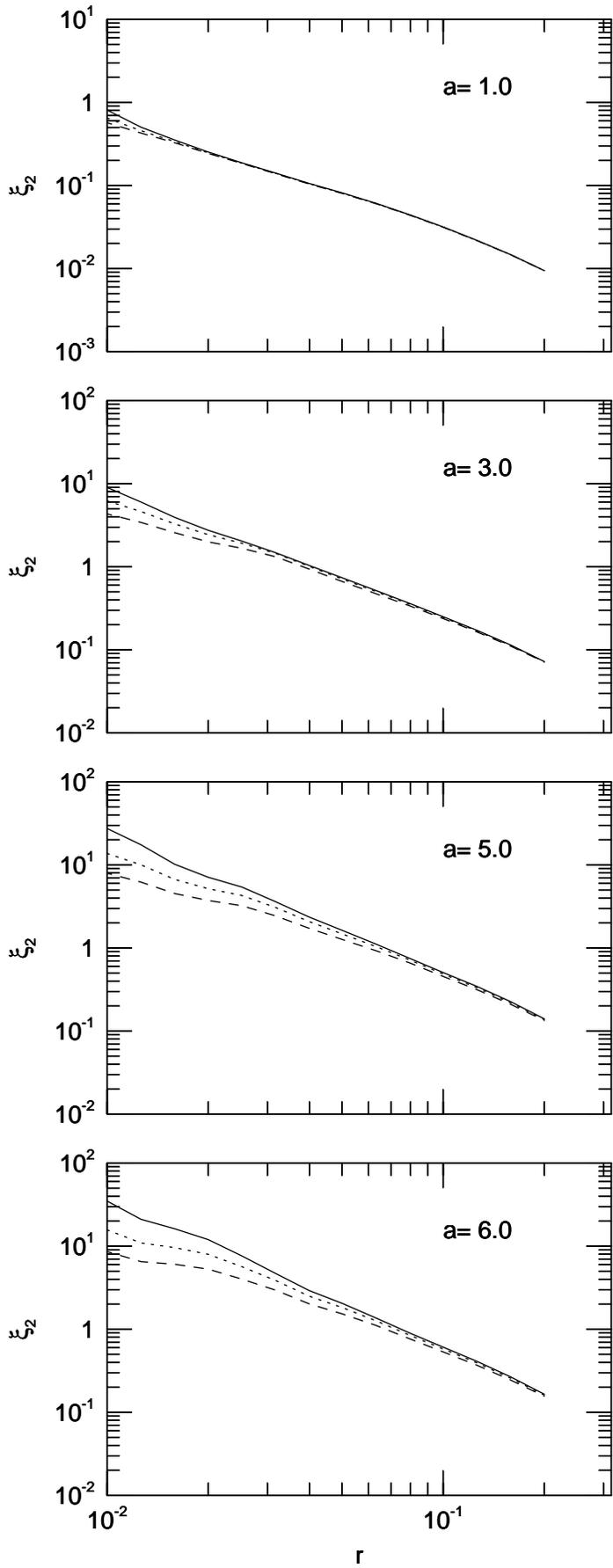

Figure 4

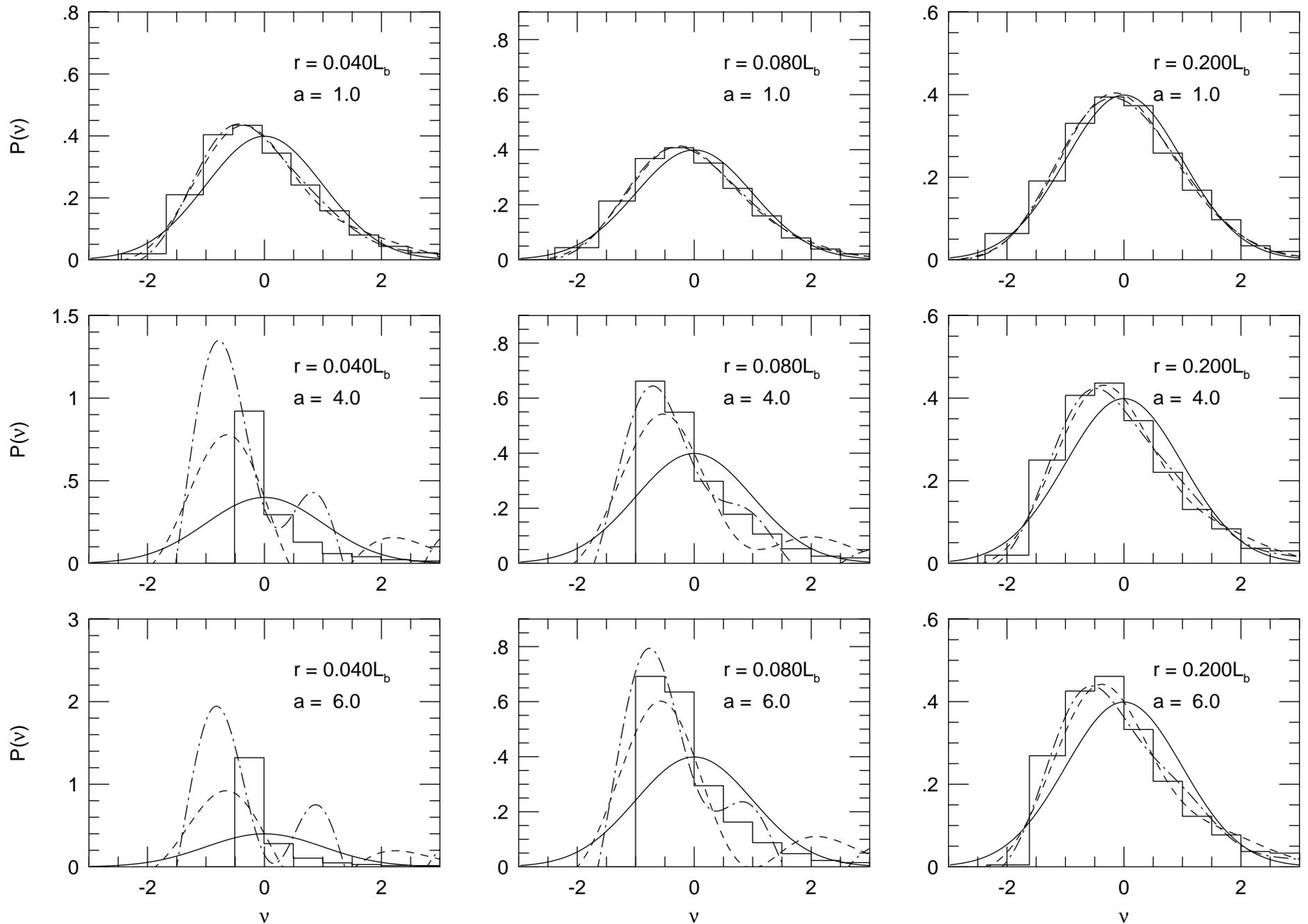

Figure 5

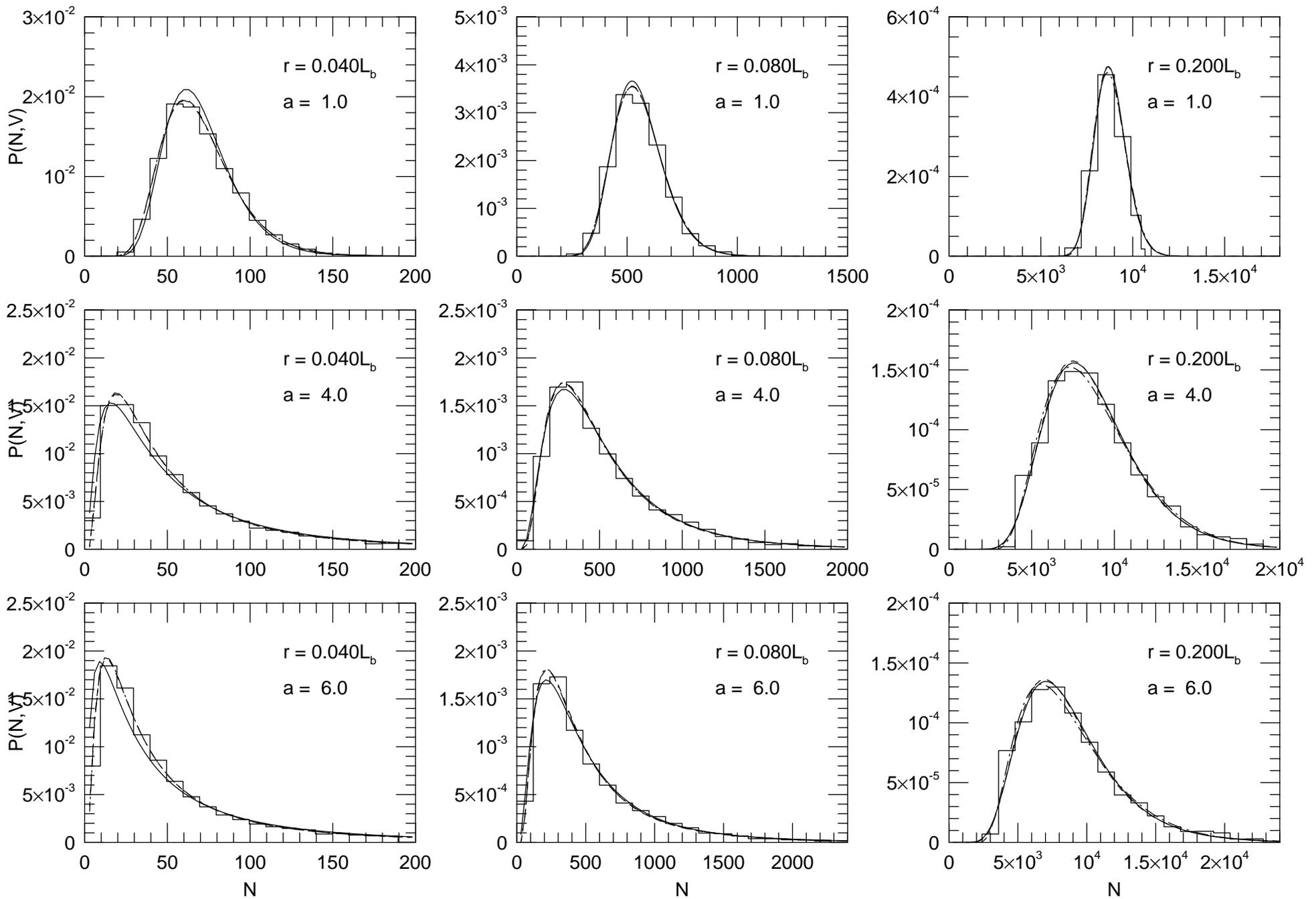

Figure 6

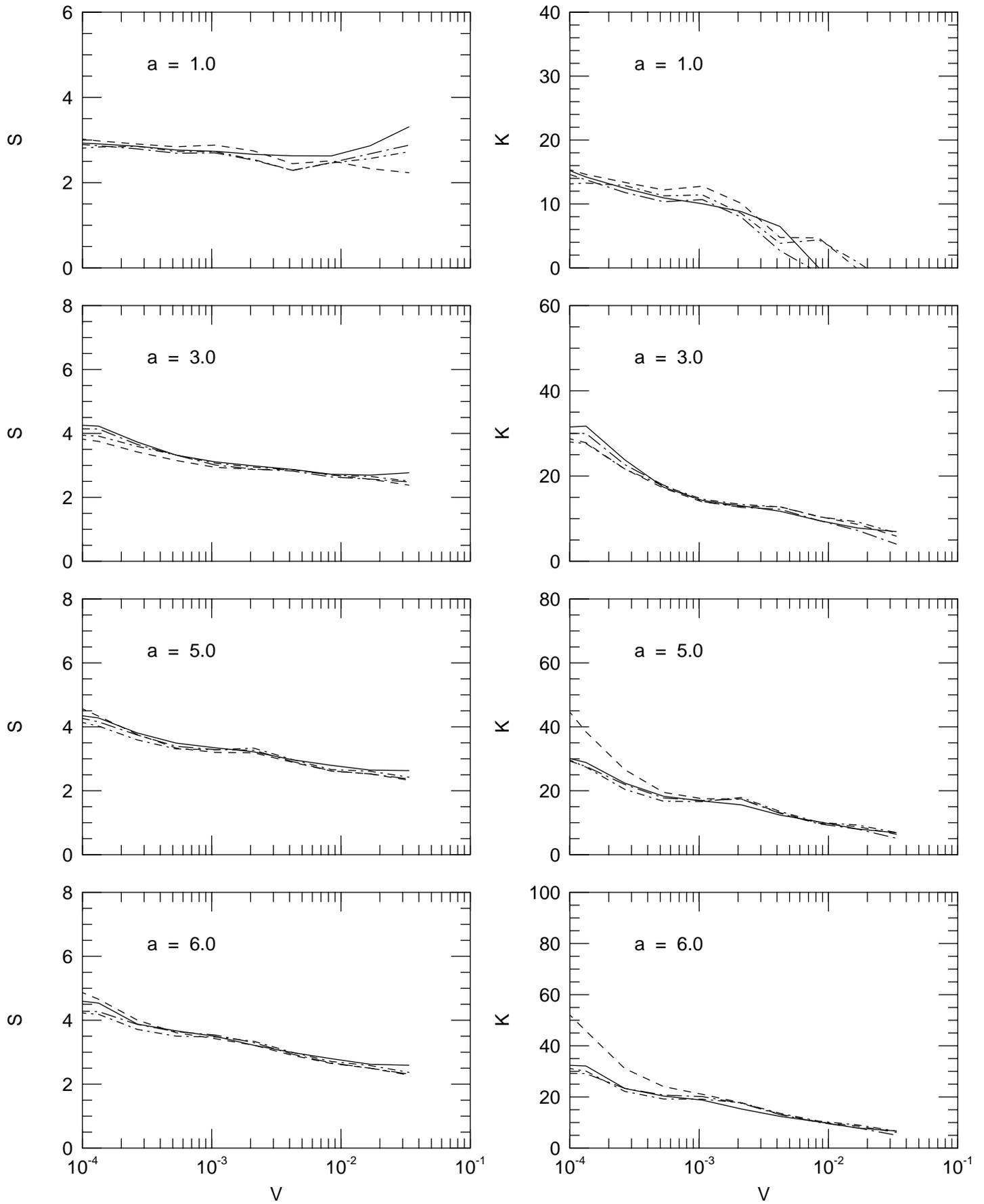

Figure 7

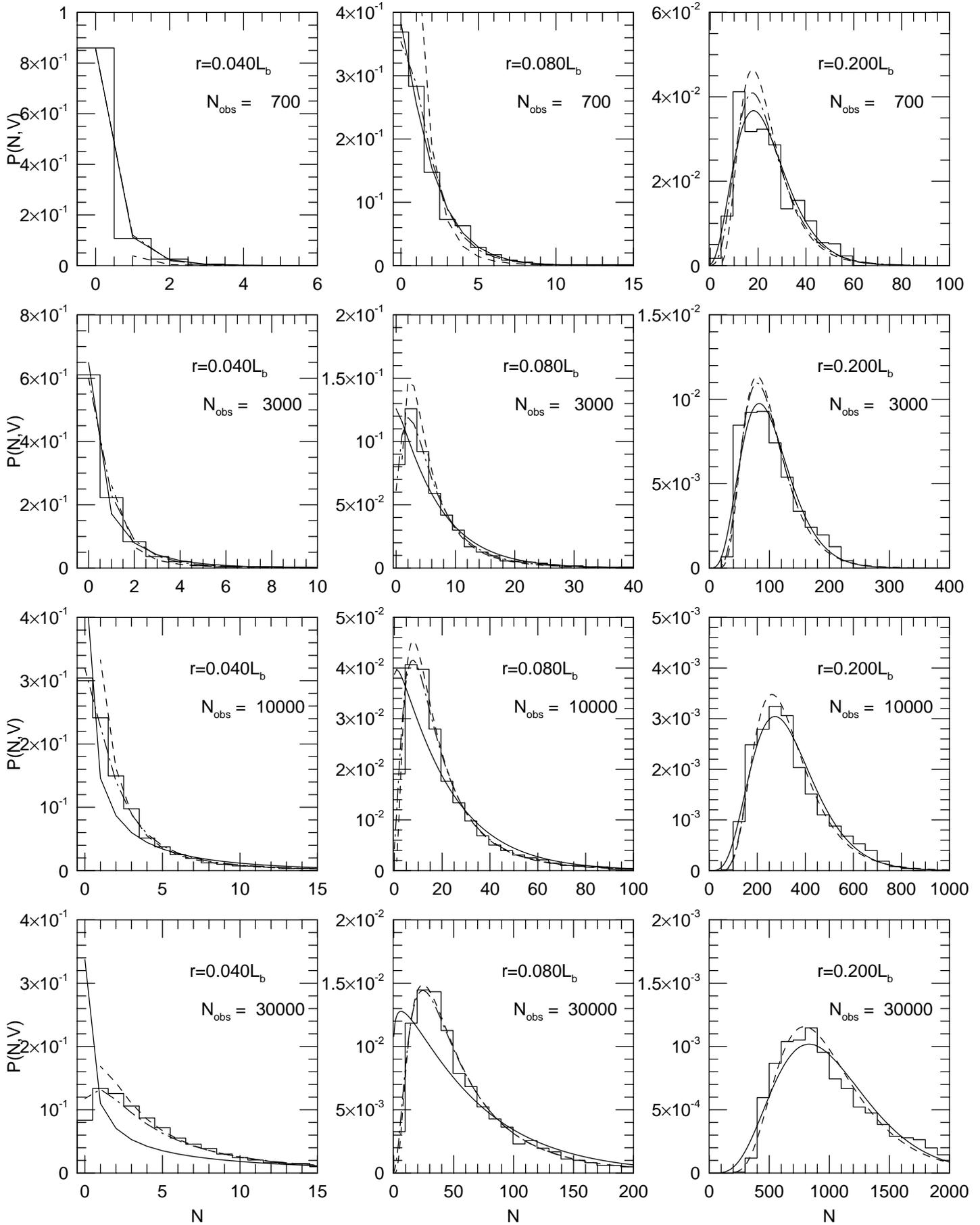